\documentclass[12pt]{article}
\input epsf.tex
\epsfclipon
\usepackage{epsfig}
\usepackage{epstopdf}
\usepackage{epsf}
\usepackage{graphicx,amsmath,amssymb}
\usepackage{epsfig,multicol}
 \usepackage{epsfig}
\usepackage{amssymb}
\usepackage{jheppub}
\def\bg{\begin{eqnarray}}
\def\nd{\end{eqnarray}}

\begin{document}
\title{Bounds on Operator Dimensions in 2D Conformal Field Theories }
\author{Joshua D. Qualls, Alfred D. Shapere}
\affiliation{Department of Physics and Astronomy, University of Kentucky, Lexington, KY 40506 USA}
\emailAdd{joshua.qualls@uky.edu, shapere@pa.uky.edu}

\abstract{
We extend the work of Hellerman \cite{hell} to derive an upper bound on the conformal dimension $\Delta_2$ of the next-to-lowest nontrival primary operator in unitary, modular-invariant two-dimensional conformal field theories without chiral primary operators, with total central charge $c_{\rm tot}>2$. The bound we find is of the same form as found by Hellerman for $\Delta_1$: $\Delta_2 \leq \frac{c_{\rm tot}}{12}+O(1)$. We obtain a similar bound on the conformal dimension $\Delta_3$, and present a method for deriving bounds on  $\Delta_n$ for any $n$, under slightly modified assumptions.  For asymptotically large $c_{\rm tot}$ and $n\lesssim \exp (\pi c/12)$, we show that  $\Delta_n \leq \frac{c_{\rm tot}}{12}+O(1)$.  This implies an asymptotic lower bound of order $\exp (\pi c_{\rm tot}/12)$ on the number of primary operators of dimension $\le c_{\rm tot}/12 + O(1)$, in the large-$c$ limit.  

In dual gravitational 
theories, this corresponds to a lower bound in the flat-space limit on the number of gravitational states without boundary excitations, of mass less than or equal to $1/4G_N$.}

\maketitle

\section{Introduction}

Conformal field theories (CFTs) describe many physical phenomena, from critical phenomena to string theory to quantum gravity through the AdS/CFT correspondence. While relatively little is known about CFTs in general, in recent years a number of constraints on their spectra and amplitudes have been obtained by means of conformal bootstrap techniques \cite{1p,2p,3p}. In \cite{4p,v21}, bounds on the conformal dimensions of operators in four-dimensional unitary CFTs were derived from the condition of crossing symmetry of the four point function, using explicit expressions for the conformal blocks obtained in \cite{5p,7p}. 
More recently, similar methods have been applied to CFTs in diverse dimensions with great success
\cite{8p,v24, v22,v23,v25, v26, v27, v28}.

In two dimensions, Hellerman \cite{hell} used modular invariance of the partition function to derive a bound on $\Delta_1$, the conformal dimension of the lowest nonvacuum primary operator, in any unitary 2D CFT with no chiral primary operators other than the identity and with left and right central charge $c, \tilde{c}>1$: 
\begin{equation}
	\Delta_1 \leq  \frac{c_{\rm tot}}{12} + 0.4736... \label{eq:hellbound}
\end{equation}
Hellerman also discussed the dual gravitational interpretation of this result, which corresponds to a bound on the lightest massive black hole state in the dual 3D gravitational theory. More recently, Friedan and Keller \cite{fried} investigated additional constraints from modular invariance systematically. Building on the work of \cite{hell}, they applied the next several differential constraints using the linear functional method and found that for finite $c_{\rm tot}$ the bound 
(\ref{eq:hellbound}) can be lowered somewhat. For large $c_{\rm tot}$, however, the bounds 
apparently 
all asymptote to $\frac{c_{\rm tot}}{12}$ as in (\ref{eq:hellbound}).

Additional assumptions on the 2D CFT lead to tighter bounds on $\Delta_1$. The paper \cite{6p} (see also \cite{9p,10p}), for example, examined 2D CFTs for which the partition function is holomorphically factorized as a function of the complex structure $\tau$ of the torus. In this class of CFT, it can be shown that the lowest primary operator is either purely left- or right-moving, and can have a weight no larger than $1 + \mbox{min}(\frac{c}{24},\frac{\tilde{c}}{24})$. Other work \cite{11p,fried} considers  a certain subclass of (2,2) SUSY CFTs that suggest a bound that goes as $\Delta_1\leq\frac{c}{24}$ for large central charge.

In this paper, we extend the arguments of \cite{hell} to derive bounds on the conformal dimensions $\Delta_2$, $\Delta_3$ using no additional assumptions.  The bounds we obtain take the same form as Hellerman's bound (\ref{eq:hellbound}), with the same asymptotic growth $c_{\rm tot}/12$. We also investigate the possibility of deriving bounds on primary operator conformal dimensions $\Delta_n$ for $n>3$. We find that in order to obtain a bound for $\Delta_4$ or higher $\Delta_n$, we need to assume a larger minimum value for $c_{\rm tot}$, which grows logarithmically with $n$.   
For large $c_{\rm tot}$ with $c_{\rm tot}\gtrsim \frac{12}\pi \log n$, we show that 
all the $\Delta_n$ obey a bound of the same form as (\ref{eq:hellbound}): 
$$
\Delta_n\le \frac{c_{\rm tot}}{12}+O(1).
$$
These bounds, satisfied for fixed $c_{\rm tot}$ by all $\Delta_n$ with $ \log n \lesssim \pi c_{\rm tot}/12$,
 collectively imply that the total number of primaries of dimension $\Delta \lesssim c_{\rm tot}/12$ grows at least exponentially with $c_{\rm tot}$ 
 $$
N(c_{\rm tot}/12) \gtrsim \exp \left( \frac {\pi c}{12}\right)
$$

In cases where the CFT has a gravitational dual, our results have implications for the spectrum of the gravitational theory through the AdS/CFT correspondence.  In the final section, we show that our lower bound on the number of primaries of dimension $\lesssim c/12$ is consistent with the entropy of a black hole of mass $1/4G_N$.

\section{Review of the bound on $\Delta_1$}

We begin by reviewing the methods and results of \cite{hell}. Consider a 2D CFT on the torus with modular parameter close to the fixed point $\tau\equiv\ (K+i\beta)/2\pi=i$, where  $\beta$ is the inverse temperature and $K$ is the thermodynamic potential for spatial momentum in the compact spatial direction $\sigma_1$. The path integral on the square torus corresponds to the thermal partition function of the CFT compactified on a circle. We can parameterize the neighborhood of this fixed point conveniently using $\tau\equiv i \mbox{ exp}(s)$. Then invariance of the partition function $Z(\tau,\bar{\tau})$ under the modular $S$-transformation $\tau \rightarrow -\frac{1}{\tau}$ can be expressed as 
\begin{equation}
Z\left(i e^s, - i e^{\bar{s}}\right) =Z(i e^{-s},-i e^{-\bar{s}})
\end{equation}
By taking derivatives of this expression with respect to $s, \bar{s}$, one obtains an infinite set of equations

\begin{equation}\left(\tau\frac{\partial}{\partial \tau}\right)^{N_{L}}  \left(\bar{\tau}\frac{\partial}{\partial \bar{\tau}}\right)^{N_{R}} Z(\tau,\tilde{\tau})\bigg|_{\tau=i}=0, \;\;N_{L} + N_{R} \text{ odd} \end{equation}
For purely imaginary complex structure $\tau = i\beta/2\pi$, this condition implies 
\begin{equation}\left(\beta\frac{\partial}{\partial \beta}\right)^{N}   Z(\beta)\bigg|_{\beta=2\pi}=0, \;\;N \text{ odd} \label{eq:twothreeish} \end{equation}

We will assume a unique vacuum and a discrete spectrum. By further assuming cluster decomposition and no chiral operators other than the stress tensor, the Virasoro structure theorem implies that the partition function  $Z(\beta)$ can be expressed as a sum over conformal families:
\begin{equation}
Z(\beta)=Z_{id}(\beta) + \sum_A Z_{A}(\beta).
\end{equation}
Here $Z_{id}(\beta)$ is the sum over states in the conformal family of the identity; $Z_{A}(\beta)$ is the sum over all states in the conformal family of the $A^{th}$ primary operator, which has conformal weights $h_A, \tilde{h}_A$ and conformal dimension $\Delta_A\equiv h_A +\tilde{h}_A $. 

Hellerman considers CFTs with $c,\tilde{c}>1$ and with no chiral operators other than the stress tensor, which implies the following explicit forms for $Z_{id}(\beta)$ and $Z_{A}(\beta)$:
\begin{equation}
Z_{id}(\tau)=q^{-\frac{c}{24}}\bar{q}^{-\frac{\tilde{c}}{24}}\prod_{m=2}^{\infty}{ (1-{q}^{m})^{-1}}\prod_{n=2}^{\infty}{(1-\bar{q}^{n})^{-1}}
\end{equation}
\begin{equation}
Z_{A}(\tau)=q^{h_A-\frac{c}{24}}\bar{q}^{\tilde{h}_A-\frac{\tilde{c}}{24}}\prod_{m=1}^{\infty}{ (1-{q}^{m})^{-1}}\prod_{n=1}^{\infty}{(1-\bar{q}^{n})^{-1}}
\end{equation}
where $q=\exp(2\pi i \tau)$.
The full partition function with $\tau = i\beta/2\pi$ is then given by the expression
\begin{equation}
Z(\beta)= M(\beta)Y(\beta)+B(\beta) \label{eq:twofiveish}
\end{equation}
with
\begin{equation}
M(\beta)\equiv \frac{\exp(-\beta\hat{E}_0)}{\eta(i\beta/2\pi)^2}
\end{equation}
and
\begin{equation}
B(\beta)\equiv M(\beta) \left(  1-\exp(-\beta) \right)^{2},
\end{equation}
where $\hat{E}_0 \equiv E_0 + \frac{1}{12} = \frac{1}{12}-\frac{c+\tilde{c}}{24}$ and $\eta$ is the Dedekind eta function. For real $\beta$, the partition function over primaries $Y(\beta)$ is
\begin{equation}
Y(\beta)=\sum_{A=1}^{\infty}e^{-\beta \Delta_{A}}.
\end{equation}

Next, Hellerman applies the differential constraints (\ref{eq:twothreeish}) to the partition function (\ref{eq:twofiveish}). To simplify the analysis, we introduce polynomials $f_p(z)$ defined by
\begin{equation}
(\beta \partial_\beta)^{p}M(\beta)Y(\beta)\bigg|_{\beta=2\pi} = (-1)^{p} \eta(i)^{-2}\text{exp}(-2\pi \hat{E}_0) \sum_{A=1}^{\infty}\text{exp}(-2\pi \Delta_A)f_p(\Delta_A+\hat{E}_0).
\end{equation}
The first few polynomials are explicitly

$$
f_0(z)=1 $$
$$f_1(z)=(2\pi z)-\frac12 $$
\begin{equation}f_2(z)=(2\pi z)^2-2(2\pi z)+\left(\frac{7}{8}+2r_{20}\right)\label{eq:fs}\end{equation}
$$f_3(z)=(2\pi z)^3-\frac{9}{2}\left(2\pi z\right)^2+\left(\frac{41}{8}+6r_{20}\right)(2\pi z)-\left( \frac{17}{16}+3r_{20}\right)$$
where 
\begin{equation*}
r_{20} \equiv \frac{\eta''(i)}{\eta(i)}\approx0.0120...
\end{equation*}
We also define the polynomials $b_p(z)$ by
\begin{equation}
(\beta \partial_\beta)^{p}B(\beta)\bigg|_{\beta=2\pi} = (-1)^{p} \eta(i)^{-2}\text{exp}(-2\pi \hat{E}_0)b_p(\hat{E}_0),
\end{equation}
Explicitly,
$$
b_0(z)=1 -2e^{-2\pi}+e^{-4\pi}
$$
\begin{equation}
b_1(z)=\left( (2\pi z)-\frac12 \right)- 2e^{-2\pi}\left((2\pi (z+1))-\frac12 \right) + e^{-4\pi}\left((2\pi (z+2))- \frac12 \right)  \label{eq:bs}
\end{equation}
$$
b_p(z)= f_p(z) - 2e^{-2\pi} f_p(z+1) + e^{-4\pi}f_p(z+2) .
$$
Using these polynomials, the equations (\ref{eq:twothreeish}) for modular invariance of $Z(\beta)$ for odd $p$ become

\begin{equation}
\sum_{A=1}^{\infty}f_p(\Delta_A+\hat{E}_0)\text{exp}(-2\pi\Delta_A)=-b_p(\hat{E}_0)  \label{eq:modeq}
\end{equation}
 
It is this expression that is used to derive an upper bound on the conformal dimension $\Delta_1$. Hellerman takes the ratio of the $p=3$ and $p=1$ expressions to get
\begin{equation}
\frac{\sum_{A=1}^{\infty}f_3(\Delta_A+\hat{E}_0)\text{exp}(-2\pi\Delta_A)}{\sum_{B=1}^{\infty}f_1(\Delta_B+\hat{E}_0)\text{exp}(-2\pi\Delta_B)}     =    \frac{b_3(\hat{E}_0)}{b_1(\hat{E}_0)}\equiv F_1.
\end{equation}
Or, upon rearrangement,
\begin{equation}
\frac{\sum_{A=1}^{\infty}   \left[ f_3(\Delta_A+\hat{E}_0)  - F_1(\hat{E}_0) f_1(\Delta_A+\hat{E}_0)    \right]        \text{exp}(-2\pi\Delta_A)}{\sum_{B=1}^{\infty}f_1(\Delta_B+\hat{E}_0)\text{exp}(-2\pi\Delta_B)}     =   0. \label{eq:hellratio}
\end{equation}

Next assume that $\Delta_1 > \Delta_1^+$, where $\Delta_1^+$ is defined as the largest root of the numerator, and proceeds to obtain a contradiction. Because $\Delta_A \geq \Delta_1$, this assumption implies that every term in both the numerator and denominator is strictly positive. Then equation (\ref{eq:hellratio})  says that a sum of positive numbers equals zero --- an impossibility. Therefore
$$
\Delta_1 \leq \Delta_1^+.
$$ 
Finally, by analyzing $\Delta_1^+$ as a function of $c_{\rm tot}$ Hellerman proves that for the given assumptions, $\Delta_1^+ \leq \frac{c_{\rm tot}}{12}+   \frac{(12-\pi) + (13\pi -12)e^{-2\pi}}{6\pi(1-e^{-2\pi})}$, implying the bound
\begin{equation}
 \Delta_1 \leq \frac{c_{\rm tot}}{12}+ 0.4736...
\end{equation}

\section{Bounds on $\Delta_2$, $\Delta_3$}

In this section, we extend the methods described above to derive bounds on primary operators of second and third-lowest dimension. In order to bound the conformal dimension $\Delta_2$, we move the $\Delta_1$ term of equation (\ref{eq:modeq}) to the RHS. We then form the ratio of the $p=3$ and $p=1$ equations to get
\begin{eqnarray}
\frac{\sum_{A=2}^{\infty}f_3(\Delta_A+\hat{E}_0)e^{-2\pi\Delta_A}}{\sum_{B=2}^{\infty}f_1(\Delta_B+\hat{E}_0)e^{-2\pi\Delta_B}} = \frac{f_3(\Delta_1+\hat{E}_0)e^{-2\pi\Delta_1}+b_3(\hat{E}_0)}{f_1(\Delta_1+\hat{E}_0) e^{-2\pi\Delta_1}+b_1(\hat{E}_0)} \equiv F_2(\Delta_1,c_{\rm tot}).  \label{eq:fdef}
\end{eqnarray}
Moving $F_2$ to the left side,  we get
\begin{equation}
\frac{\sum_{A=2}^{\infty} \left[f_3(\Delta_A+\hat{E}_0)-  f_1(\Delta_A+\hat{E}_0  ) F_2\right]\text{exp}(-2\pi\Delta_A)}{\sum_{B=2}^{\infty}f_1(\Delta_B+\hat{E}_0)\text{exp}(-2\pi\Delta_B)} =0  \label{eq:mine}
\end{equation}
In Appendix $A$, we prove that $F_2$ is finite and nonzero for $c,\tilde{c}>1$ and $\Delta_1$ in the allowed range and thus our derivations will carry through without issue.

Before proceeding, we make some definitions. Define $\Delta^+_{f_p}$ to be the largest root of $f_p(\Delta+\hat{E}_0)$ viewed as a polynomial in $\Delta.$ The bracketed expression in the numerator is a polynomial cubic in $\Delta_2$; we denote it by $P_2(\Delta_2)$, and define the largest root of $P_2$ to be $\Delta^+_2(c_{\rm tot},\Delta_1)$, where $\hat{E}_0$ dependence has been replaced by $c_{\rm tot}$. 
	
We now assume that $\Delta_2>\text{ max}(\Delta^+_{f_1}, \Delta^+_2)$ and attempt to obtain a contradiction. From our explicit polynomial expressions, we see that the leading coefficients of both $f_1$ and $f_3$ are positive. Thus both $P_2(\Delta_2)>0$ and $f_1(\Delta_2+\hat{E}_0)>0$ for $\Delta_2>\text{ max}(\Delta^+_{f_1}, \Delta^+_2)$. Because $\Delta_n\geq\Delta_2$ for all $n>2$, we also have $P_2(\Delta_n)>0$ and $f_1(\Delta_n+\hat{E}_0)>0$ for $\Delta_2>\text{ max}(\Delta^+_{f_1}, \Delta^+_2)$.  Thus every term in both the numerator and denominator of the left side of equation (\ref{eq:mine}) is positive for $\Delta_2>\text{ max}(\Delta^+_{f_1},\Delta^+_2)$. The left side thus can not be equal to zero, and we have a contradiction. We have thus derived a bound on the conformal dimension $\Delta_2$:
\begin{equation}
\Delta_2\leq \text{ max}(\Delta^+_{f_1},\Delta^+_2). \label{eq:bbound}
\end{equation}

From the explicit form of $f_1(\Delta+\hat{E}_0$) in (2.12), we see that 
\begin{equation}
\Delta^+_{f_1} = \frac{c_{\rm tot}}{24}+\frac{(3-\pi)}{12\pi}. \label{eq:delfp}
\end{equation}
We will spend the next section trying to simplify our bound by deriving a manageable expression for $\Delta^+_2$.

\subsection*{Asymptotic expansion for large central charge}

We begin by considering the limit of large positive total central charge $c_{\rm tot}$. In the limit $c_{\rm tot}\rightarrow\infty$, it is easy to see that $\Delta^+_2$ is proportional to $c_{\rm tot}$, plus corrections of order $c^0_{\rm tot}$. We thus expand $\Delta^+_2$ as a series at large central charge:
\begin{equation}
\Delta^+_2\equiv\sum_{a=-1}^{\infty}d_{-a}(\Delta_1)\left( \frac{c_{\rm tot}}{24}\right)^{-a}. \label{eq:serdef}
\end{equation}
By definition $\Delta^+_2$ satisfies
\begin{equation*}
P_2(\Delta^+_2)=0
\end{equation*}
and is the largest real value with that property.
Substituting equation (\ref{eq:serdef}) into the explicit form of $P_2(\Delta^+_2)=0$, the equation to leading order in $c_{\rm tot}$ is:

\begin{equation}
\frac{4(d_1-1)^2\pi^2}{24^2}-\frac{\pi^2}{12^2}=0.
\end{equation}
The solution $d_1=2$ gives the largest root $\Delta_2^+$,
\begin{equation}
\Delta^+_2=\frac{c_{\rm tot}}{12}+d_0(\Delta_1)+O(c_{\rm tot}^{-1}).
\end{equation}
Note how this compares to the value $\Delta^+_{f_1}$ in equation (\ref{eq:delfp}). Since we are taking the maximum of these two quantities, the true upper bound on $\Delta_2$ will generically be given by $\Delta_2^+$.

To determine $d_0$ we expand $P_2$ to the next order in $c_{\rm tot}$. Quoting the result obtained in Appendix $B$, we find that the largest term possible at this order is given by $d_0 \approx 0.4736...$---the same bound to this order as for $\Delta_1$ in \cite{hell}. Thus for large enough central charge $c_{\rm tot}$, we can always bound the conformal dimension $\Delta_2$ using the expression
\begin{equation}
\Delta_2 \leq \frac{c_{\rm tot}}{12}+ 0.4736... + O(c^{-1}_{\rm tot}).
\end{equation}

An absolute bound on $\Delta_2$ can be obtained numerically. We seek a linear bound of the form $\Delta_2 \leq \frac{c_{\rm tot}}{12} + D_1$, where $D_1$ is a numerical constant independent of $\Delta_1$. In order for this bound to universal, we need to find $D_1$ so that the inequality is valid for all possible values of $\Delta_1$ and all $c_{\rm tot}> 2$. This can be done by explicitly solving the cubic polynomial $P_2$ (in terms of radicals of exponentials) and maximizing the expression $\Delta^+_2 - \frac{c_{\rm tot}}{12}$ for $c_{\rm tot}>2$ and $ 0<\Delta_1 \leq \frac{c_{\rm tot}}{12}+ 0.4736..$. This function attains a global maximum $D_1 \approx 0.5338...$ (for $c_{\rm tot} \approx 2,\,\, \Delta_1 \approx 0.2717...).$ Therefore 
\begin{equation}
\Delta_2 \leq \frac{c_{\rm tot}}{12} + 0.5338...
\end{equation}

\subsection*{Proof and numerical bound for $\Delta_3$}
Now that we have obtained a bound on $\Delta_2$, it is natural to extend our arguments to primary operators of higher dimension. A necessary condition for our 
arguments to work for $\Delta_n$ is that $F_{n}$, defined as
\begin{equation}
F_{n}(\hat{E}_0,\Delta_1,\cdots,\Delta_{n-1}) \equiv \frac{\sum_{i=1}^{n-1} f_3(\Delta_i+\hat{E}_0)\text{exp}(-2\pi\Delta_i)+b_3(\hat{E}_0)}{\sum_{i=1}^{n-1}f_1(\Delta_i+\hat{E}_0)\text{exp}(-2\pi\Delta_i)+b_1(\hat{E}_0)},  \label{eq:fndef}
\end{equation}
be well-defined for all relevant values of its arguments. We prove in Appendix A that $F_3$ is well-defined for $c_{\rm tot}> 2$ and thus that there will be no issues. We can thus proceed with another proof by contradiction. The result is that 
\begin{equation}
\Delta_3 \leq \text{max}(\Delta^+_{f_1},\Delta_3^+), \label{eq:3bound}
\end{equation}  
where $\Delta^+_{f_1}$ is the expression (\ref{eq:delfp}) from above and $\Delta_3^+$ is the largest real root of the polynomial
\begin{equation}
P_3(\Delta_3) \equiv f_3(\Delta_3+\hat{E}_0)-f_1(\Delta_3+\hat{E}_0) F_3.
\end{equation}

At large central charge, one easily finds that $\Delta_3^+ \approx \frac{c_{\rm tot}}{12}$. Maximizing the expression $\Delta_3^+ - \frac{c_{\rm tot}}{12}$ numerically as a function of $\Delta_1$ and $\Delta_2$ subject to the contraints $0<\Delta_1\leq \frac{c_{\rm tot}}{12}+ .4736...$, $0<\Delta_2\leq \frac{c_{\rm tot}}{12}+D_1$, and $c_{\rm tot}> 2$ gives the constant $D_2 = 0.8795...$ and the linear bound
\begin{equation}
\Delta_3 \leq \frac{c_{\rm tot}}{12}+ 0.8795...
\end{equation}
We observe that the values of $\Delta_{1,2}$ that maximize $\Delta_3^+$ are degenerate: $\Delta_1=\Delta_2$. 

\section{Bounds on $\Delta_n$}

It should be clear by this point how to extend the proof to higher conformal dimensions. Assuming that the expression (\ref{eq:fndef}) is defined and nonvanishing in the appropriate range, we can proceed as above to obtain a bound
\begin{equation}
\Delta_n \leq \text{max}(\Delta^+_{f_1},\Delta_n^+), 
\end{equation}  
where $\Delta^+_{f_1}$ is given by (\ref{eq:delfp}) and $\Delta_n^+$ is the largest real root of the polynomial
\begin{equation}
P_n(\Delta_n) \equiv f_3(\Delta_n+\hat{E}_0)-f_1(\Delta_n+\hat{E}_0) F_{n}(c_{\rm tot}, \Delta_1,...,\Delta_{n-1}) \label{eq:pn}
\end{equation}
and is thus a function of $c_{\rm tot}, \Delta_1, \cdots, \Delta_{n-1}$.

The leading terms in the polynomial with largest root $\Delta_n^+$ are independent of $n$; therefore the expansion of $\Delta_n^+$ at asymptotically large central charge again goes as $\frac{c_{\rm tot}}{12}$. Thus it seems reasonable to expect a bound of the same form as before: 
\begin{equation}
\Delta_n \leq \Delta_n^+ <  \frac{c_{\rm tot}}{12} + O(1).
\end{equation}
However, there is a potential problem with this argument. For the bounds on $\Delta_2$ and $\Delta_3$, we proved in the Appendices that the functions $F_2$ and $F_3$ were positive and well-defined for the relevant ranges of our parameters. This is not the case beginning with the expression $F_4$. The denominator of $F_4$ vanishes when the total central charge equals
\begin{equation} 
c_{D4} = \frac{ 2 [\sum_{i=1}^{3}(-12\pi\Delta_i-\pi+3 )  e^{-2\pi\Delta_i}  -\pi + 3 +  (26\pi - 6) e^{-2\pi} +( 3 -25\pi) e^{-4\pi} ] }  {\pi ( \sum_{i=1}^{3}-e^{-2\pi\Delta_i} -1 +2 e^{-2\pi}-e^{-4\pi} ) }
\end{equation}
As before, we extremize this expression over the appropriate ranges of its variables ($0< \Delta_1\leq c_{\rm tot}/12 + 0.4736...$, etc.). The largest value of the total central charge for which the denominator of $F_4$ vanishes is given by 
\begin{equation}
c^+_{D4} = 2.3450...
\end{equation}
Applying the same analysis to the numerator of $F_4$, we find that the largest value of the total central charge causing it to vanish is
\begin{equation}
c^+_{N4} = 1.5113...
\end{equation}
Thus for $2 < c_{\rm tot} < c_{D4}^+$,we cannot use these specific methods to set a bound on $\Delta_4$; there is a moduli space where our parameters can fundamentally change the polynomial $P_n$.

The resolution to this issue is straightforward; we will further restrict the allowed values for the total central charge to $c_{\rm tot}>\mbox{max}(c^+_{D4},c^+_{N4})$. Allowing $c_{\rm tot}$ to range all the way down to 2.3450... would require an infinite constant in the bound (4.3); however, a small additional restriction on the range to $c_{\rm tot}\ge 2.5$ leads to the bound
\begin{equation}
\Delta_4 \leq \frac{c_{\rm tot}}{12} + 1.0795...  \label{eq:del4bound}\end{equation}
Further restricting $c_{\rm tot}$ gives a tighter bound; for example, $c_{\rm tot} \geq 3$ gives $\Delta_4 \leq \frac{c_{\rm tot}}{12} + 0.6740...$
Similar results can be derived for arbitrary $\Delta_n$ using the methods described here. We note that, as before, the values of $\Delta_{1,2,3}$ that saturate the bound (\ref{eq:del4bound}) are degenerate: $\Delta_1=\Delta_2=\Delta_3$.

For larger values of $n$, it can be shown that $c_{Dn}^+ > c_{Nn}^+$ and thus we need only restrict $c_{\rm tot}>c_{Dn}^+$.
We can analytically solve for the value of the central charge $c_{Dn}$ which causes the denominator of (\ref{eq:fdef}) to vanish. The explicit form is a complicated function of $\Delta_1,...,\Delta_{n-1}$ in terms of Lambert W functions; we provide details in Appendix C. We also show there that when we maximize $c_{Dn}$ over all of its arguments, it goes for large $n$ as  
\begin{equation}
c^+_{Dn}\approx \frac{12}{\pi} W_0[(n-1)] \sim \frac{12}{\pi}\log (n),  \label{eq:dtor}
\end{equation} 
where $W_0$ is the primary branch of the Lambert-$W$ function. 
Therefore, if we require 
\begin{equation}
\log{n} \lesssim \frac{\pi c_{\rm tot}}{12} + O(1) \label{eq:nbound},
\end{equation}
then $F_n$ will be finite and nonzero. Then an analysis similar to before gives a bound
\begin{equation}
\Delta_n \leq \frac{c_{\rm tot}}{12} + O(1)\label{eq:deltanbound}.
\end{equation}

The $O(1)$ term in expression (\ref{eq:deltanbound}) means $O(1)$ in $c_{\rm tot}$--- these subleading terms could have dependence on $n$ that contributes to leading order. For example, if the $O(1)$ term goes as $\log(n)$, then by equation (\ref{eq:nbound}) we could have contributions as large as $O(c_{\rm tot})$. Additionally, the specific $O(1)$ term will depend on how we restrict the total central charge. In Appendix D, we show that by considering
\begin{equation}
n \ll e^{\pi c_{\rm tot}/6} + O(1) \label{eq:truenbound},
\end{equation}
we can derive a bound on $\Delta_n$ for asymptotically large $c_{\rm tot}$ going as
\begin{equation}
\Delta_n \leq \frac{c_{\rm tot}}{12} + O(1). \label{eq:finalbound}
\end{equation}
In the limit (\ref{eq:truenbound}), the $O(1)$ term will be 0.4736... and additional corrections will be $O(n c_{\rm tot} e^{\pi c_{\rm tot} / 6})$. We are already assuming eq. (\ref{eq:nbound}), so the inequality (\ref{eq:truenbound}) necessarily follows.

\section{Gravitational Interpretation}

Our results have implications for gravity in $2+1$ dimensions through the AdS/CFT correspondence. In the case of AdS$_3$/CFT$_2$, the matching between the central charge of the CFT and the cosmological constant is given by the identification
\begin{equation}
c+\tilde{c}=\frac{3L}{G_N},
\end{equation}
where $L$ is the AdS radius and $G_N$ is Newton's constant. Following \cite{witten}, we  also match the spectrum of massive states with the spectrum of primary operators---the usual AdS/CFT dictionary gives the correspondence 
\begin{equation}
E^{(rest)}=\frac{\Delta}{L},
\end{equation}
where $E^{(rest)}$ is the rest energy of an object in the bulk of AdS and $\Delta$ is the conformal dimension of the corresponding boundary operator.

We can interpret our bounds as saying that the dual gravitational theory, when it exists, must have massive states in the bulk (without boundary excitations) with rest energies $M_n=\Delta_n/L$ satisfying
\begin{equation}
M_n\leq M_n^+\equiv\frac{1}{L}\Delta^+_n |_{c_{\rm tot}=\frac{3L}{G_N}}.
\end{equation} 
Using our asymptotic bound (\ref{eq:deltanbound}), this inequality becomes
\begin{equation}
M_n\leq \frac{1}{4G_N}+\frac{D_n}{L}
\end{equation}
where $D_n$ is an $O(1)$ or smaller constant and $n$ is constrained in the same appropriate manner. In the flat-space limit $L\rightarrow\infty$, this inequality becomes
$$
M_n \leq  \frac{1}{4G_N}
$$ 
It is interesting that $ \frac{1}{4G_N}$ is close to the mass of the lightest  BTZ black hole in pure (2+1)-D gravity.  

Since $n$ can be of exponentially large order in $c$ according to eq.(\ref{eq:nbound}), 
this inequality indicates a high density of gravitational microstates of mass  $\le 1/4 G_N$.  Indeed, the logarithm of the number $N$ of such states should be at least equal to the upper bound on $\log(n)$ of  eq.(\ref{eq:nbound}), 
\begin{equation}
\label{eq:logn}
\log{N} \ge \frac{\pi c_{\rm tot}}{12} + O(1)= \frac {\pi L}{4 G_N}+ O(1)
\end{equation}
The density of these states should be strongly peaked at the upper limit of the mass range, so this may be interpreted as a lower bound on the entropy of a spinless (2+1)-D black hole of mass $1/4G_N$.
Indeed, the actual entropy of a spinless black hole of this mass is  \cite{19p,20p}
$$
S=\frac{\pi c_{\rm tot}}6 = \frac {\pi L}{2 G_N}
$$
which  is indeed consistent with the bound (\ref{eq:logn}).

\section*{Acknowledgments}

The authors would like to thank several individuals for helpful discussions and feedback, including Benjamin Braun, Diptarka Das, Sumit Das, Michael Eides, Tom Hartman, and especially Simeon Hellerman. This work is partially supported by a University of Kentucky fellowship and by NSF $\#0855614$ and $\#1214341$.

\appendix
\section{Behavior of $F_2, F_3$}
\setcounter{section}{1}

In this appendix, we prove that the functions $F_2, F_3$ are defined for all relevant values of our parameters. The function $F_2$ is given by 
\begin{equation}
F_2 \equiv \frac{f_3(\Delta_1+\hat{E}_0)\text{exp}(-2\pi\Delta_1)+b_3(\hat{E}_0)}{f_1(\Delta_1+\hat{E}_0)\text{exp}(-2\pi\Delta_1)+b_1(\hat{E}_0)}
\end{equation}
and the polynomials $f_i$ and $b_i$ are given in equations (\ref{eq:fs}) and (\ref{eq:bs}). By inspection, we see that $F_2$ will only become undefined if the denominator equals zero. The value of the central charge when $f_1(\Delta_1+\hat{E}_0)\text{exp}(-2\pi\Delta_1)+b_1(\hat{E}_0)=0$ is
\begin{equation}
c^+_{D2} = \frac{2}{\pi} \frac{ (12\pi\Delta_1 +\pi - 3) e^{-2\pi\Delta_1}+\pi-3-26\pi e^{-2\pi} + 6 e^{-2\pi}+ 25\pi e^{-4\pi}-3 e^{-4\pi}}
{e^{-2\pi\Delta_1}+1-2 e^{-2\pi}+e^{-4\pi}}
\end{equation}

The maximum possible value that $c^+_{D2}$ can take for $\Delta_1 > 0$ is $c^+_{D2}= 1.0868...$ This value of the total central charge, however, is outside of the assumed range $c_{\rm tot}>2$. 
Therefore the function $F_2$ is defined for all relevant values of $c_{\rm tot}, \Delta_1$. 

Our proof can also run into problems if $F_2$ is vanishing for any values of our parameter space. The condition for vanishing $F_2$ is
\begin{equation}
f_3(\Delta_1+\hat{E}_0)\text{exp}(-2\pi\Delta_1)+b_3(\hat{E}_0)=0.
\end{equation}
We once again solve for the total central charge satisfying this equation and label it. This expression can be maximized numerically; it has a maximum value given by $c^+_{N2} = 0.9632...$. This value of the total central charge is also outside of the relevant range $c_{\rm tot} > 2$. 
Therefore the function $F_2$ is well-defined and non-vanishing--in fact, positive-- for all relevant values of $c_{\rm tot}, \Delta_1$, and our proof by contradiction will be valid.

A similar analysis applies to the function $F_3$  given by 
\begin{equation}
F_3 \equiv \frac{\sum_{i=1}^{2} f_3(\Delta_i+\hat{E}_0)\text{exp}(-2\pi\Delta_i)+b_3(\hat{E}_0)}{\sum_{i=1}^{2}f_1(\Delta_i+\hat{E}_0)\text{exp}(-2\pi\Delta_i)+b_1(\hat{E}_0)}.
\end{equation}
Once again, we are interested in where this function vanishes or becomes undefined. This can be studied by solving for values of the central charge at which either the numerator or denominator vanishes. These solutions will be labeled as $c_{N3}$ and $c_{D3}$; they are functions of $\Delta_1$ and $\Delta_2$. We maximize $c_{N3}$ and $c_{D3}$ over the allowed range of $\Delta_1, \Delta_2$ and find
\begin{equation}
c^+_{N3}\approx 1.3929...
\end{equation}
and
\begin{equation}
c^+_{D3}\approx 1.8022...
\end{equation}
These values of the central charge, however, are again outside of the relevant range for $c_{\rm tot}$, since we have restricted our work to $c_{\rm tot} > 2$. 
Therefore the function $F_3$ is defined and positive for all relevant values of $c_{\rm tot}, \Delta_1,$ and $\Delta_2$.

\section{The $O(1)$ term in $\Delta_2^+$}
\setcounter{section}{2}

In this appendix, we calculate the $O(1)$ term in the expansion of the largest root $\Delta_2^+$ of polynomial $P_2$ for asymptotically large total central charge. In the body of the text, we reasoned that the leading coefficient in the large-$c_{\rm tot}$ expansion of $\Delta_2^+$,
$$
\Delta^+_2\equiv\sum_{a=-1}^{\infty}d_{-a}(\Delta_1)\left( \frac{c_{\rm tot}}{24}\right)^{-a},
$$
is $d_{1}=2$. Expanding $P_2(\Delta_2^+)=0$ to  next order in $c_{\rm tot}$, we find the expression
\begin{equation*}
\frac{-1}{e^{-2\pi\Delta_1}+(1 - e^{-2\pi })^2}    \frac{\pi}{18}  (-\pi e^{-2\pi\Delta_1} -6\pi d_0 e^{-2\pi\Delta_1} -13\pi e^{-4\pi} +14\pi e^{-2\pi}  -6\pi d_0- \pi -24e^{-2\pi}
\end{equation*}
\begin{equation*}
+12e^{-4\pi} +12e^{-2\pi\Delta_1} +12 -6\pi d_0 e^{-4\pi} +12\pi d_0 e^{-2\pi} -6\pi\Delta_1 e^{-2\pi*\Delta_1})=0.
\end{equation*}
Solving for $d_0$ gives us
\begin{equation*}
d_0(\Delta_1)=\frac{-(-14\pi e^{-2\pi}+13\pi e^{-4\pi}+\pi-12e^{-2\pi \Delta_1}+24e^{-2\pi} -12e^{-4\pi} +6\pi\Delta_1 e^{-2\pi\Delta_1}+\pi e^{-2\pi\Delta_1}-12)}{6\pi(e^{-2\pi\Delta_1}+(1-e^{-2\pi})^2)}
\end{equation*}

To keep our bound universal we should take the maximum possible value of this function. This occurs as $\Delta_1 \rightarrow \infty$ meaning $d_0(\Delta_1)\rightarrow  0.4736...$---the same constant appearing in the bound on $\Delta_1$. Thus for large enough central charge $c_{\rm tot}$, we can always bound the conformal dimension $\Delta_2$ using the expression
\begin{equation*}
\Delta_2 \leq \frac{c_{\rm tot}}{12}+ 0.4736...
\end{equation*}

\section{Condition on $n$, $c_{\rm tot}$}
\setcounter{section}{3}

Here we will sketch the proof of the condition on $c_{\rm tot}$ given by equation (\ref{eq:dtor}). We begin with the condition that the denominator of $F_{n}$ vanishes
\begin{equation}
\sum_{A=1}^{n-1}f_1\left(\Delta_A + \frac{1}{12} - \frac{c_{Dn}^+}{24}\right)e^{-2\pi \Delta_A} - b_1\left(\frac{1}{12} - \frac{c_{Dn}^+}{24}\right) = 0.    \label{eq:dzero}
\end{equation}
Upon expansion, this can be rearranged to give
$$
\frac{\pi c_{Dn}^+}{12}\left( (1-e^{-2\pi})^2 + \sum_{A=1}^{n-1} e^{-2\pi \Delta_A}  \right)      =$$
$$
\sum_{A=1}^{n-1}2\pi \Delta_A e^{-2\pi \Delta_A}  + \left(\frac{\pi}{6} - \frac12 \right) \left( (1-e^{-2\pi})^2 + \sum_{A=1}^{n-1} e^{-2\pi \Delta_A}  \right) -4\pi e^{-2\pi}(1-e^{-2\pi}).
$$
Dividing through by the parenthetical expression on the LHS gives an expression for $c_D$.

We wish to consider total central charge larger than $c_{Dn}$ in equation (\ref{eq:dzero}). We maximize $c_{Dn}$ by differentiating with respect to $\Delta_i$ to find critical points. It can be shown that $c_{Dn}$ attains a local maximum when 
\begin{equation}
\label{eq:deltas}
\Delta_1 = \cdots = \Delta_{n-1} = \frac{1}{2\pi}W_0[A(n-1)] + \frac{1}{2\pi}-\frac{2}{e^{2\pi}-1},
\end{equation}
$$
A\equiv        
\frac{  \mbox{exp}\left(( -\frac{1-(1+4\pi) e^{-2\pi}}{1-e^{-2\pi}}     \right)  } {1-2e^{-2\pi}+e^{-4\pi}} = 0.3780...
\vspace{0.1in}
$$
For small $n$ we have numerically confirmed that this is a global maximum, and we assume that this continues to be the case for all $n$. In Appendix D we show that this is the case given appropriate assumptions.
Substituting (\ref{eq:deltas}) into equation (\ref{eq:dzero}) gives a complicated expression involving the Lambert-$W$ function, defined by
$$
z = W_0(z) e^{W_0(z)} \;\;\;\Rightarrow  \;\;\;e^{-W_0(z)} = \frac{W_0(z)}{z}.
$$
After some algebra, we find the expression
\begin{equation}
c_D^{+} = \frac{12}{\pi} \left(   W_0[A(n-1)] + C_1 		\right),  \label{eq:truecbound}
\end{equation}
$$
C_1 \equiv  -\frac{4\pi}{e^{2\pi}-1} + \frac{\pi}{6} - \frac12.
$$

We consider central charge such that $c_{\rm tot}>c_D^{+}$, and use the fact that $W_0(z) \approx \log(z)$ plus $O(\log(\log(z)))$ corrections. Then for large $n$ only the first term on the RHS of (\ref{eq:truecbound}) will survive, and we deduce
$$
c_{Dn}^+ \approx \frac{12}{\pi}  W_0[A(n-1)] \sim \frac{12}{\pi} \log(n)
$$
as in eq. (4.8)

\section{Derivation for $\Delta_n$}
\setcounter{section}{4}

Here we will provide a derivation of the bound (\ref{eq:finalbound}). We define $x\equiv 2\pi (\Delta_1 + \hat{E}_0) -\frac32$ in order to depress the cubic polynomial $P_n$ of eq. (\ref{eq:pn}) to
\begin{eqnarray} 
P_n(x) &=& x^3 + \hat{C}_1(\hat{E}_0) x + \hat{C}_0(\hat{E}_0) x,  \nonumber \\
\hat{C}_1(\hat{E}_0) &\equiv& \hat{C}_0(\hat{E}_0) -\frac32, \\
\hat{C}_0(\hat{E}_0) &\equiv& -F_{n} +6r_{20}-\frac18 . \nonumber
\end{eqnarray}
It is known \cite{13p} that the largest real root of a depressed cubic obeys the inequality
\begin{eqnarray}
x^+ &\leq& \sqrt{\frac{4|\hat{C}_1|}{3}} \cos\left(\frac{\phi}{3} + \frac{2\pi k}{3}   \right)  \label{eq:whit} \\
&=& \frac{2}{\sqrt{3}} \sqrt{\bigg|  F_{n}  - 6 r_2 + \frac{13}{8} \bigg| }  \cos\left(\frac{\phi}{3} + \frac{2\pi k}{3}   \right), \nonumber  
\end{eqnarray}
where $|\cos(\phi)|\equiv \sqrt{\frac{-27 \hat{C}_0^2}{4 \hat{C}_1^3}}$ and $k=0,1,2$.

The difference between the bound on $\Delta_1$ and the bound on $\Delta_n$ is the presence of factors of $n$ in terms multiplying $F_n$. Therefore we will first consider what limits suppress this additional dependence. It can be shown by explicit computation that the function $F_{n}$ (\ref{eq:fndef}) has a maximum when $\Delta_1,...,\Delta_{n-1}$ are degenerate. As we will soon see explicitly (though somewhat apparent from eq. (\ref{eq:whit})), maximizing $F_{n}$ maximizes the bound on $\Delta_n$. Thus we need to maximize $F_{n}$ as a function of $\Delta_1$.  Differentiating and solving to leading order in $c_{\rm tot}$ for the critical point $\Delta_1^{max}$ gives that $\Delta_1^{max} \sim \frac{c_{\rm tot}}{12}$ plus subleading corrections. 

From the definition of $F_n$, we see that

it contains terms depending on $n$ as large as $n c_{\rm tot}^3 e^{-2\pi\Delta_1} \sim n c_{\rm tot}^3 e^{-\pi c_{\rm tot}/6}$. We therefore impose the condition 
\begin{equation}
n  \ll c_{\rm tot}^{-3}  e^{\pi c_{\rm tot}/6} . \label{eq:slynt}
\end{equation}
In order for $F_n$ to be nonvanishing and finite in the case of large $n$, we are already restricting ourselves to the case 
\begin{equation}
\log(n) < \frac{\pi c_{\rm tot}}{12}.
\end{equation}
Thus we will have no issues suppressing these $n$-dependent terms in $F_n$.

Although the argument for a bound going as $c_{\rm tot}/12$ follows immediately, we will continue algebraic manipulations in order justify previous statements. In the limit (\ref{eq:slynt}), $F_n$ will be of the form
\begin{eqnarray}
F_{n} &\approx& \frac{a_3 c_{\rm tot}^3 + a_2 c_{\rm tot}^2+a_1 c_{\rm tot}+a_0}{\bar{a}_1 c_{\rm tot}+\bar{a}_0}    \nonumber \\
&= & c_{\rm tot}^2  \frac{a_3}{\bar{a}_1} \left( 1+O(c_{\rm tot}^{-1}) \right)  \label{eq:gammaexp} .
\end{eqnarray}
The $a_i$ and $\bar{a}_j$ are obtained from eq. (\ref{eq:fndef}) evaluated at $n=0$, except that $a_0, \bar{a}_0$ have corrections much smaller than $O(1)$. Because $\hat{C}_0$ and $\hat{C}_1$ are just $F_n$ plus constants, in the limit we consider they will be of the same form as above, with $a_1$ and $a_0$ replaced by different constants; that is, both $\hat{C}_0$ and $\hat{C}_1$ grow asymptotically like $c_{\rm tot}^2$.

We now turn our attention to the $\cos\phi$ terms in eq. (\ref{eq:whit}). From the definition of $\cos \phi$ and considerations of the preceding paragraph, the leading behavior of $\cos(\phi)$ is $O(c_{\rm tot}^{-1})$. By the series expansion of arccosine, we then have $\phi \approx \pm \frac{\pi}{2} + O(c_{\rm tot}^{-1})$. This in turn implies 
$$\mbox{max}\left[\cos\left(\frac{\phi}{3} + \frac{2\pi k}{3}   \right) \right] = \frac{\sqrt{3}}{{2}} + O(c_{\rm tot}^{-1})$$
plus subleading corrections. Then to leading order eq. (\ref{eq:whit}) becomes 
\begin{equation}
x^+ \leq \sqrt{|F_{n}|} , \label{eq:xf}
\end{equation}
plus subleading corrections. Given eq. (\ref{eq:gammaexp}), the leading term eq. (\ref{eq:whit}) is
\begin{equation}
x^+ \leq \frac{\pi c_{\rm tot}}{12} + O(1).
\end{equation}
Finally, the definition of $x^+$ gives the result
\begin{equation}
\Delta_n^+ \leq \frac{c_{\rm tot}}{12} + O(1).
\end{equation}

As previously stated, this 
result could have been argued once we placed the appropriate restrictions on $n$ and $c_{\rm tot}$. 
The details of the preceding paragraph can also be used to justify our assertion that  the above bound on $\Delta_n$ is attained at the minimum allowed central charge.  To see this, consider maximizing the expression $(\Delta_n^+-\frac{c_{\rm tot}}{12})$. Inspection of eqs. (\ref{eq:gammaexp}) and (\ref{eq:xf}) shows that the subleading $c_{\rm tot}$ dependence comes from the expansion of the square root of the ratio of polynomials in powers of $c_{\rm tot}^{-1}$. Analysis of this square root shows that it is monotonically decreasing over the allowed range of $c_{\rm tot}$. So to maximize the square root, we should let  $c_{\rm tot}$ take its smallest allowed value. 

We have noted that the allowed range for $c_{\rm tot}$ should not extend all the way to  the pole at $c_D^+$  (\ref{eq:truecbound}); otherwise the constant appearing in (\ref{eq:deltanplus}) will diverge. It turns out this restriction is not necessary.  By considering the ratio of the $p=3,5$ constraints, one can show that a bound of form (\ref{eq:deltanplus})  is implied, with a smaller $O(1)$ term that remains finite all the way  down to the value of $c_D^+$ found above.  Thus, eq. (\ref{eq:truecbound}) is indeed a true lower bound on the number of primary operators satisfying $\Delta \lesssim {c_{\rm tot}/12}$.

\end{document}